\documentclass[twocolumn,amsmath,amssymbprl]{revtex4}

\usepackage{graphicx}
\usepackage{dcolumn}
\usepackage{bm}

\def\gtrsim{\buildrel > \over {_{\sim}}}

\begin{document}


\title{Cosmic-ray electron injection from the ionization of nuclei}

\author{Giovanni Morlino}
\affiliation{INAF/Osservatorio Astrofisico di Arcetri, 
Largo E. Fermi, 5 - 50125 Firenze (Italy)}

\date{\today}

\begin{abstract}
We show that the secondary electrons ejected from the ionization of heavy
ions can be injected into the acceleration process that occurs at
supernova remnant shocks. This electron injection mechanism works since
ions are ionized during the acceleration when they move already with
relativistic speed, just like ejected electrons do. Using the abundances of
heavy nuclei measured in cosmic rays at Earth,  we estimate the
electron/proton ratio at the source to be $\sim 10^{-4}$, big enough to
account for the nonthermal synchrotron emission  observed in young SNRs. We
also show that the ionization process can limit the maximum energy that
heavy ions can reach.
\end{abstract}
\pacs{Valid PACS appear here}

\maketitle

Supernova Remnant (SNR) are believed to be the primary sources of Cosmic
Rays (CR) in the Galaxy. The theory of Diffusive Shock Acceleration (DSA)
applied to SNR blast waves propagating in the interstellar medium
provides the most comprehensive framework for the explanation of the CR
spectrum measured at Earth. Nevertheless some aspects of DSA still remain
unsolved and one of the most fundamental issues concerns how electrons can
be injected into the acceleration process. The presence of accelerated
electrons is a well established fact, deduced from direct observations of
young SNRs, where both Radio and X-ray emission are interpreted as
synchrotron radiation of highly relativistic electrons \cite{reynolds}.

The injection problem is related to the shock dynamics which is dominated
by protons (and maybe also by heavier ions). The shock layer is expected
to be few thermal proton gyroradii thick, which means that particles need
have a few times the mean thermal proton momentum in order to cross the
shock and undergo DSA. The injection condition can be write as $p>p_{\rm
inj} \equiv \xi p_{p,th}$, where $p_{p,th}= \sqrt{2 m_p \,K_B T_{p,2}}$ is
the typical downstream thermal proton momentum and $\xi$ contains the
complex shock microphysics \cite{kang}. According to theoretical estimates
\cite{BGV} and observational constraints \cite{morlino}, $\xi$ is in the
range 2-4.
The injection condition can be easily fulfilled for suprathermal
protons which reside in the highest energy tail of the Maxwellian
distribution. On the other hand the same condition, applied to electrons,
is satisfied only in the relativistic regime with a minimum Lorentz factor
$\gamma_{\rm inj}\approx 3$-$30$ for a typical $T_{p,2}$ in the range
$10^6$-$10^8$ K. It is therefore hard to imagine how electrons can come
from the thermal component: if we assume that electrons upstream of the
shock thermalize downstream their bulk kinetic energy, i.e. $K_B T_{e,2}=
\frac{1}{2}m_e u_{\rm shock}^2$, then the mean thermal electron momentum is
$p_{e,th}= (m_e/m_p)\, p_{p,th}$. Even assuming some mechanism able to
equilibrate quickly electrons and protons at the same temperature, the mean
electron momentum rises only up to $p_{e,th}= \sqrt{m_e/m_p}\, p_{p,th}$. 

Most proposed solutions for this injection problem involve some kind of
pre-acceleration mechanism able to accelerate electrons from thermal
energies up to mildly relativistic energies. Some studies predict that
electrons can be effectively pre-accelerated by electrostatic waves
generated in the shock layer \cite{galeev,levinson}. 
These mechanisms are difficult to study analytically, because their
understanding requires the knowledge of the complex microphysics that
regulates collisionless shocks. A better way to investigate them is through
particle-in-cell \cite{amano} and Monte Carlo simulations \cite{baring},
which, unfortunately, are still not able to provide firm conclusions on
the electron injection efficiency.

In this paper we show that the ionization of heavy nuclei during 
acceleration can inject a number of mildly relativistic electrons large
enough to account for the synchrotron radiation observed in young SNRs. 
In fact, nuclei heavier than hydrogen in the ISM where SNR shocks
propagate, are never fully ionized simply because the typical ISM
temperature is not large enough, being of the order of $10^4-10^5$ K. This
statement is also supported by the presence of Balmer-dominated filaments
observed in several young remnants \cite{sollerman}, showing that even the
hydrogen in the ISM is not fully ionized.

Once the shock encounters a partially ionized atom, the atom can start DSA
in the same way protons do. We note that a correct computation of the
injection of heavy ions involves the knowledge of the initial charge and
the aggregate state of atoms and their downstream temperature, which are
very difficult to predict. We neglect such complications and we assume that
the injection of heavy elements occurs, simply because they are observed in
the CR spectrum. The key point we want to stress here is that, once the
acceleration begins, atoms are not stripped immediately, because the
ionization time turns out to be large enough to allow them to reach
relativistic energies before complete ionization. When atoms move
relativistically, the ionization can occur either via Coulomb collisions or
via photoionization. In both cases, the mean kinetic energy of ejected
electrons, measured in the ion rest frame, is negligible with respect to
the electron mass energy. Hence ejected electrons move, in the plasma rest
frame, approximately along the same direction and with the same speed of
the parent atoms. In this case, the momentum of ejected electrons can
easily exceed $p_{\rm inj}$. In order to prove this statement we start
comparing the acceleration with the ionization time.

Let us consider a single partially ionized species $N$, with mean charge
$Z_{\rm eff}$, atomic charge $Z$ and mass $m_N= 2Z m_p$. For simplicity we
compute the acceleration time in the framework of linear shock acceleration
theory, i.e. neglecting the dynamical role of accelerated particles, and
for a plane shock geometry. If a particle with momentum $p$ diffuses with a
diffusion coefficient $D(p)$, the well known expression for the
acceleration time is \cite{drury}: 
\begin{equation}
 t_{acc}(p)= \frac{3}{u_1-u_2} \left( \frac{D_1(p)}{u_1} +
 \frac{D_2(p)}{u_2} \right) \,,
 \label{eq:t_acc1}
\end{equation}
where $u$ is the plasma speed in the shock rest frame, and the subscript 1
(2) refers to the upstream (downstream) quantities (note that $u_{\rm
shock} = u_1$). The downstream and upstream plasma speeds are related 
through the compression factor, $u_2= u_1/r$. We limit our considerations
to strong shocks, which have compression factor $r=4$, and we assume Bohm
diffusion coefficient, i.e. $D_B=r_L \beta c/3$, where $\beta c$ is the
particle speed and $r_L= pc/(Z_{\rm eff} eB)$ is the Larmor radius. The
turbulent magnetic field responsible for particle diffusion is assumed to
be compressed downstream according to $B_2= r\,B_1$. Even if this relation
applies only for the magnetic component parallel to the shock plane,
choosing a different compression rule does not affect our main results.
Applying previous assumptions, Eq.~(\ref{eq:t_acc1}) becomes:
\begin{equation}
 t_{acc}(\gamma)= 1.7\, \left(\gamma -\gamma^{-1} \right) \, B_{\mu G}^{-1}
 \, u_8^{-2} \left( Z/Z_{\rm eff} \right) \,{\rm yr} \,,
 \label{eq:t_acc2}
\end{equation}
where $\gamma$ is the particle Lorentz factor. Here $B_{\mu G}$ is the
upstream magnetic field expressed in $\mu$G and the shock speed is
$u_1= u_8\, 10^8 {\rm cm/s}$. In order to compute the energy reached by
particles when ionization occurs, we compare Eq.~(\ref{eq:t_acc2}) with
the ionization time. As already mentioned, ionization can occur either
via Coulomb collisions with thermal particles or via photoionization by
background photons. Whether the former process dominates on the latter
depends on the ratio between the thermal particle and the ionizing photon
densities.

We consider first the role of collisions. A full treatment of collisional
ionization is hard because of the complex cross section involved and it is
beyond the scope of this work; our purpose can be achieved using the
following approximation. Let us consider the process in the rest
frame of the target atom, which is bombarded with point-like charged
particles (in our case protons or electrons) with kinetic energy $E_{\rm
kin}$. The classical ionization cross section has a maximum when
$E_{\rm kin}$ is twice the ionization energy, $I$, and the value is:
\begin{equation}
 \sigma_{\rm coll}^{\rm max} = \pi a_0^2 N_e\, I_{\rm Ryd}^{-2} \,,
 \label{eq:Coulomb}
\end{equation}
where $a_0$ is the Bohr radius, $I_{\rm Ryd}$ is the ionization potential
expressed in Rydberg units and $N_e$ is the number of electrons in the
considered atomic orbital. For $E_{\rm kin}>2 I$ the cross section
decreases like $E_{\rm kin}^{-2}$ and reaches a minimum when relativistic
effects become important, while, in the full relativistic regime
$\sigma_{\rm coll} \propto \log(E_{\rm kin})$ (see e.g. \cite{inokuti}). 
Hence we can use Eq.~(\ref{eq:Coulomb}) as a good upper limit for the
collisional cross section in a wide range of incident particle energy.

Accelerated ions collide mainly with thermal protons and electrons, whose
densities are assumed to be equal, $n_e=n_p$. The total ionization time is
the average between the upstream and downstream contribution, weighted for
the respective residence time, i.e. $\tau_{\rm
coll}=(t_1+t_2)(t_1/\tau_{\rm coll,1}+t_2/\tau_{\rm coll,2})^{-1}$, 
where $t_i= 4D_i/cu_i$ \cite{drury}. The final result is:
\begin{equation}
 \tau_{\rm coll} \approx \left(c \, \sigma_{\rm coll}^{\rm max}\, 
    n_1 (1+r)\right)^{-1} 
  = 0.0024\,I_{\rm Ryd}^2 \, n_1^{-1} \, {\rm yr} \,,
 \label{eq:coll_time}
\end{equation}
where $n_1$ is the upstream proton density in $\rm cm^{-3}$.
Now, equating $\tau_{\rm coll}$ with the acceleration time in
Eq.~(\ref{eq:t_acc2}), we get the value of the Lorentz factor, $\gamma_{\rm
coll}$ that ions have when the collisional ionization occurs. Using typical
parameters for a young SNR, the result reads:
\begin{equation}
 \gamma_{\rm coll} \simeq 9 \,
 \left( \frac{0.1\, \rm cm^{-3}}{n_1} \right)
 \left( \frac{B_1}{25\, \mu G} \right)
 \left( \frac{u_1}{5000\, \rm km/s} \right)^2
 I_{\rm Ryd}^2 .
 \label{eq:gamma_coll}
\end{equation}
An upstream magnetic field around $25 \mu$G is expected if magnetic
amplification occurs, but the condition  $\gamma_{\rm coll}>\gamma_{\rm
inj}$ can be easily fulfilled even for magnetic field as low as the mean
galactic value, i.e. $\sim 5 \mu$G.

In the atom rest frame ejected electrons can have kinetic energy ranging
from 0 up to $E_{\rm kin}-I$. But, due to the long-range nature of Coulomb
interaction, events with a small momentum transfer are highly favored and
the majority of ejected electrons have kinetic energy $E\ll m_e c^2$
\cite{inokuti}. As a consequence when parent atoms move
relativistically with respect to the plasma rest frame, ejected electrons
move approximately in the same direction and with the same Lorentz factor
computed in Eq.~(\ref{eq:gamma_coll}). The same is true when the electron
is ejected by photoionization.

Photoionization can occur only when atoms collide with photons whose energy
is larger than the ionization potential $I$. Atoms moving relativistically
see a distribution of photons peaked in the forward direction of motion,
with a mean photon energy $\epsilon'= \gamma \epsilon$, where $\epsilon$ is
the photon energy in the plasma rest frame. In order to estimate the
photoionization time we adopt the simplest approximation for the $K$-shell
cross section \cite{heitler}, i.e.:
\begin{equation}
 \sigma_{\rm ph}(\epsilon') = 64\,\alpha^{-3}\sigma_T Z^{-2}
  \left( I/\epsilon'\right)^{7/2} \,
 \label{eq:sigma_ph}
\end{equation}
where $\sigma_T$ is the Thompson cross section and $\alpha$ is the fine
structure constant. The factor $Z^{-2}$ is due to the nuclear charge
dependence of $K$-type orbitals dimension. In order to get the full
photoionization time we need to integrate over the total photon energy
spectrum, i.e.:
\begin{equation}
 \tau_{\rm ph}^{-1}(\gamma)= \int d\epsilon\, 
 \frac{dn_{ph}(\epsilon)}{d\epsilon} \,c\,\sigma_{ph}(\gamma \epsilon) \,,
 \label{eq:ph_time}
\end{equation}
where $dn_{ph}/d\epsilon$ is the photon spectrum as seen in the plasma rest
frame. The relevant ionizing photons are only those with energy $\epsilon
\simeq I/\gamma$ (measured in the plasma frame) because the photoionization
cross section rapidly decreases with increasing photon energy.
The corresponding numerical value of photoionization time is:
\begin{equation}
 \tau_{\rm ph}(\gamma) \simeq 0.01\, Z^2 \, \left( n_{\rm ph}(I/\gamma)
 /{\rm cm^{-3}} \right)^{-1} \, {\rm yr} \,.
 \label{eq:Photoion_time}
\end{equation}
Comparing $\tau_{\rm ph}$ with Eq.~(\ref{eq:coll_time}) for the last inner
orbital (which has $I_{\rm Ryd}= Z^2$) we see that photoionization
generally dominates over the collisional ionization for heavy ions, namely
when $n_{\rm ph}\gtrsim 4 n_1 Z^{-2}$.

\begin{figure}
\begin{center}
{\includegraphics[angle=0,width=0.8\linewidth]{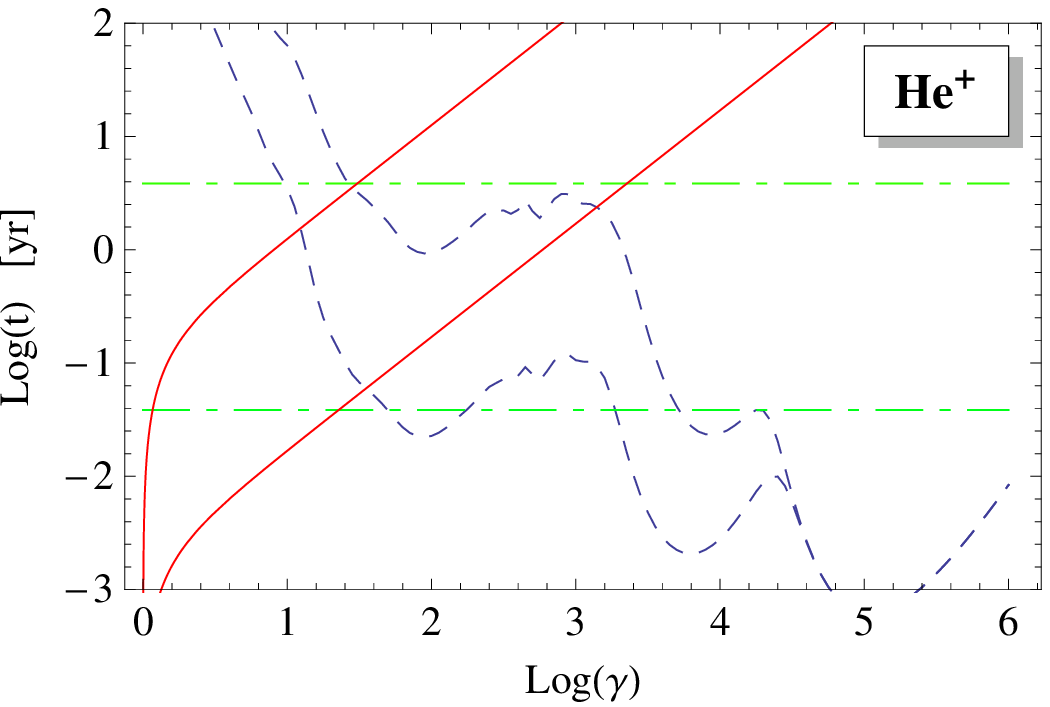}}
{\includegraphics[angle=0,width=0.8\linewidth]{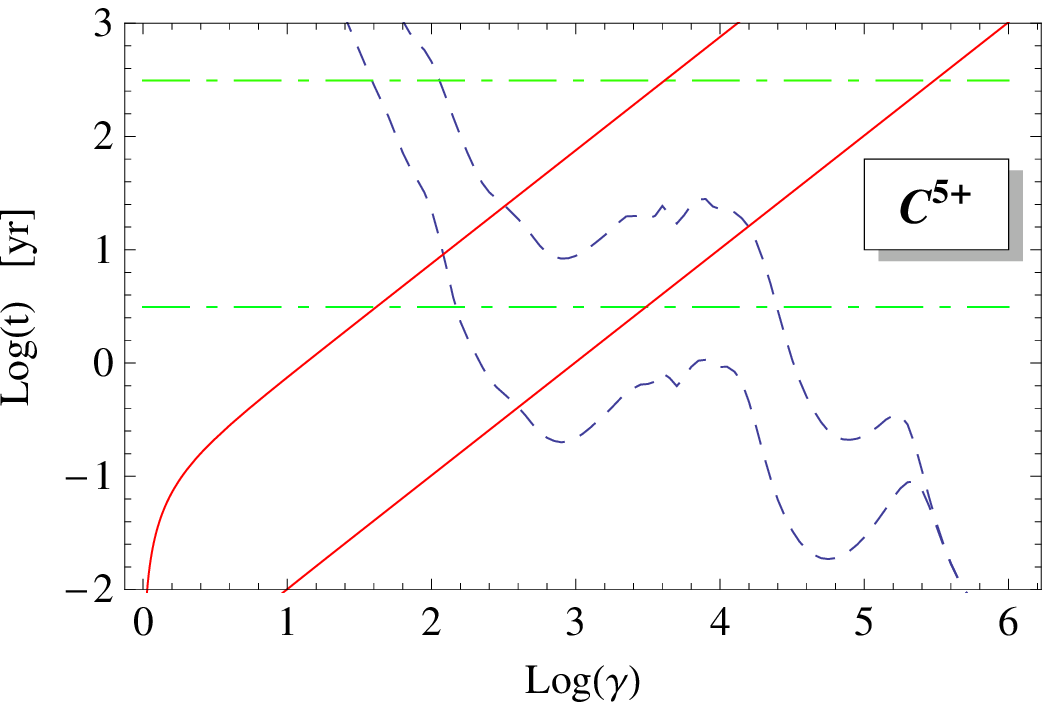}}
\caption{Comparison between acceleration (solid lines), photoionization
(dashed) and collisional ionization time (dot-dashed) as functions of the
ion's Lorentz factor. The top and lower panels show the results for the
hydrogen-like ions He$^{+}$ and C$^{5+}$, respectively. For each time two
curves are shown, rapresenting two different set of parameters, as
explained in the text.}
\label{fig:time}
\end{center}
\end{figure}

In Fig.~\ref{fig:time} we compare the acceleration time with both the
collisional and the photoionization time. The two panels show the case of
ionization for two hydrogen-like ions, He$^{+}$ and C$^{5+}$, so that
$Z_{\rm eff}=Z-1$. Each characteristic time is shown for two different
choices of the parameters:
$t_{\rm acc}$, plotted with solid lines, is shown for $u_1=3000$ km/s,
$B_1=3 \mu$G (upper line) and for $u_1=10^4$ km/s, $B_1=20 \mu$G (lower
line); $\tau_{\rm coll}$ (dot-dashed) is shown for $n_1= 0.01$ (upper
line) and $1\,\rm cm^{-3}$ (lower line); finally $\tau_{\rm ph}$ (dashed
line) is computed according to Eq.~(\ref{eq:ph_time}), using the Galactic
interstellar radiation field (ISRF) plus the cosmic microwave background.
We use the ISRF as calculated in \cite{porter}, which includes the photons
produced by stars and the infrared radiation resulting from the stellar
light reprocessed by Galactic dust. In Fig.~\ref{fig:time} the lower dashed
line is computed using the ISRF in the Galactic center, while the upper
dashed line corresponds to a location in the Galactic plane, 12 kpc far
away from the Galactic center \cite{porter}. We neglect the high energy
radiation coming from the remnant itself because the typical number density
of the X-ray photons is negligible and does not exceed $\sim 10^{-7} \rm
ph/cm^3$. 
Fig.~\ref{fig:time} shows that the most relevant contribution to 
photoionization comes from the optical photons, which produces the first
dip present of the dashed curves.
The value of $\gamma$ where $\tau_{\rm coll}$ and $\tau_{\rm ph}$ intersect
$t_{\rm acc}$ , identifies the Lorentz factor of ejected electrons. From
the upper panel we see that even electrons from He$^+$ can be ejected
with $\gamma>10$ and can easily undergo DSA.

A remarkable consequence of the ionization process is that, under
appropriate circumstances, heavy elements reach a maximum energy lower than
$Z \times E_{\rm max}^{\rm proton}$, the value predicted by the shock
acceleration theory. This because they cannot be completely ionized in a
time less than the Sedov time of a typical SNS. In fact the photoionization
of the last inner shell, due to photons with energy $\epsilon$, occurs only
when $\gamma>Z^2{\rm Ryd}/\epsilon$. The acceleration time required to
reach such a Lorentz factor is $t_{\rm acc}= 2.4 (Z/26)^2 (\epsilon/
{\rm eV})^{-1} (B_1/20{\rm \mu G})^{-1} (u/5000 {\rm km/s})^{-2} {\rm yr}$.
If the optical photon density is low enough, the photoionization is
dominated by IR photons ($\epsilon \sim 10^{-3}{\rm eV}$), and $t_{\rm
acc}$ can be longer than the Sedov time. 

Now we estimate whether the number of electrons injected into the
accelerator is large enough to produce the observed synchrotron
emission. In the literature the number of accelerated electrons is usually
compared with that of protons: DSA operates in the same way for both kind
of particles, hence a proportionality relation between their distribution
functions is usually assumed, i.e. $f_e(p)= K_{ep} f_p(p)$ (valid in the
energy range where electron losses can be neglected).
It is worth stressing that here we are only interested in the
electron/proton ratio in young SNRs, and not to the $K_{ep}$ measured
in the CR spectrum at Earth. These two quantities could be different
because the latter is the sum of the contribution coming from all sources
integrated during the source age, and also reflects transport to Earth and
losses in transport (expecially radiative losses for electrons).

The value of $K_{ep}$ in the source strongly depends on the assumption for
the magnetic field strength in the region where electrons radiate and, in
the context of the DSA theory, it can be determined for those SNRs where
both nonthermal X-ray and TeV radiation are observed. Two possible
scenarios have been proposed \cite{aharonian, berezhko, morlino}. In the
first one electrons produce both the X-ray and the TeV components via the
synchrotron emission and the inverse Compton effect, respectively; this
scenario requires a downstream magnetic field around $20\mu$G, and
$K_{ep}\sim 10^{-2}-10^{-3}$.  The second scenario assumes that the number
of accelerated protons is large enough to explain the TeV emission as due
to the decay of neutral pions produced in hadronic collisions. In this case
the DSA requires a magnetic field strength of few hundreds $\mu$G and
$K_{ep} \sim 10^{-4}-10^{-5}$. Such a large magnetic field is consistently
predicted by the theory as a result of the magnetic amplification
mechanisms which operate when a strong CR current is present. In the
following we show that injection via ionization can account for the second
scenario, i.e. the one with efficient CRs production.

In order to get the electron spectrum, $f_e(p)$, we need to solve first
the transport equations for all partially ionized species which take part
in the acceleration process and release electrons. Then we can use the ions
distribution functions as source terms for the electron transport equation.
For the sake of simplicity let us consider the simplest case where the
acceleration involves only one hydrogen-like species which inject
one electron per atom, $N^+\rightarrow N^{++} + e^{-}$, as could be the
case for He$^+$. The acceleration of all the three components can be
described using the well known transport equation \cite{drury} but addind a
``decay term'' to take into account the ionization process, i.e.:
\begin{equation}
 u \frac{\partial f_{i}}{\partial x} = D(p) \frac{\partial^{2}
 f_{i}}{\partial x^{2}} + 
 \frac{1}{3}\frac{du}{dx}p\frac{\partial f_{i}}{\partial p} +
 Q_{i} - S_i \,,
 \label{eq:trans}
\end{equation}
where the index $i= e, N^+, N^{++}$ identifies the species. $Q_i$ is the
source term while $S_i$ is the ``decay term'' due to the ionization. We
assume that the injection of ions $N^+$ occurs only at the shock position
and at a fixed momentum $p_{\rm inj}$, hence $Q_{N^+}(x,p)= K \,
\delta(p-p_{\rm inj}) \, \delta(x)$, where the normalization constant $K$
is determined by the total number of ions injected per time unit. The decay
term is $S_{N^+}=f_{N^+}(x,p)/ \tau_{\rm ion}(p)$, where the total 
ionization time is  $\tau_{\rm ion}= (\tau_{\rm coll}^{-1}+\tau_{\rm
ph}^{-1})^{-1}$.  For electrons and $N^{++}$ the decay terms vanish
while the injection terms can be approximated as follows:
\begin{equation}
 Q_{i}(x,p)= \int_p^\infty d^3p' \frac{f_{N^+}(x,p')}{\tau_{\rm ion}(p)}
 \, \delta^{(3)}(p-\xi_i p') \;, i=e, N^{++}\,. 
 \label{eq:Q_el}
\end{equation}
Because both $N^{++}$ and $e^{-}$ move approximately with the same Lorentz
factor of $N^+$, we can set $\xi_i= 1$ for $N^{++}$ and $\xi_i= m_e/m_N$
for electrons. In the case of linear shock acceleration theory,
Eq.~(\ref{eq:trans}) can be solved using standard techniques \cite{drury}
and we will show the detailed procedure in a future paper.
We define $p_0$ as the momentum value where the ionization time for $N^+$
is comparable to its diffusion time $\sqrt{4 D/u^2}$. It is easy to show
that $f_{N^{++}}(p_N)$ and $f_e(p_e)$ both become a power law $\propto
p^{-s}$ for $p_N > p_0$ and $p_e > p_0\, \times \frac{m_e}{m_N}$,
respectively. The index $s$ is only a function of the compression factor,
$s=3r/(r-1)$, and $s \rightarrow 4$ when the shock is strong. In the limit
$p\gg p_0$ the ratio between electrons and ions has the following
expression:
\begin{equation}
 K_{eN} \equiv \lim_{p \gg p_0} \frac{f_e(p)}{f_{N^{++}}(p)} = 
 \frac{Z}{2Z-1} \left( \frac{m_e}{m_N} \right)^{s-3} \,.
 \label{eq:K_ep}
\end{equation}
Here the factor $Z/(2Z-1)$ is due to the different diffusion coefficient
for electrons and ions, while the ratio $m_e/m_N$ is due to the different
momentum they have when the ionization occurs. In order to give
an approximated estimate for $K_{ep}$ we need to multiply
Eq.~(\ref{eq:K_ep}) by the total number of ejected electrons, i.e.
$(Z-Z_{\rm eff})$, and sum over all atomic species present in the
accelerator, i.e.  $K_{ep}\simeq \sum_N K_{Np}\,(Z_N-Z_{N,\rm eff})
\,K_{eN}$,  where $K_{Np}$ are the abundances of ions measured at the
source in the range of energy where the ionization occurs. 
Even if the values of $K_{Np}$ are widely unknown, we can estimate it
using the abundances measured at Earth and adding a correction factor to
compensate for propagation effects, namely the fact that particles with
different $Z$ diffuse in a different way. The diffusion time in the Galaxy
is usually assumed to be $\tau_{\rm diff}\propto (p/Z)^{-\delta}$, with
$\delta\approx 0.3-0.6$ (see \cite{blasi} for a review on recent CR
experiments). If $K_{Np,0}$ is the ion/proton ratio measured at
Earth, than the same quantity measured at the source is $K_{Np}=
K_{Np,0}Z_N^{-\delta}$. Hence the final expression for the electron/proton
ratio at the source is: 
$K_{ep}\simeq \sum_N K_{Np,0}\, Z_N^{-\delta}\,(Z_N-Z_{N,\rm eff}) \,
K_{eN}$. 

Using the abundances of nuclei measured at 1 TeV \cite{wiebel} and
assuming that $Z/2$ is the number of electrons effectively injected by
each species, we have $K_{ep}\sim 10^{-4}$.  Remarkably this number is
exactly the order of magnitude required to explain the X-ray emission in
the context of efficient CR acceleration. We note that the present estimate
is based on linear shock acceleration theory, while a correct treatment
requires the inclusion of non linear effects.

The author is grateful to P. Blasi, E. Amato, D. Caprioli and R. Bandiera
for valuable discussions and comments, and continuous collaboration.

\end{document}